# New perspectives on transient stability between grid-following and grid-forming VSCs

Kaizhe Zhang, Mei Su, *Member, IEEE*, Yao Sun, *Member, IEEE* and Zhangjie Liu, *Member, IEEE*

*Abstract*—**The grid-following and grid-forming controls in voltage-source converters are considered as different operation modes and the synchronization mechanism of them are studied separately. In this article, the intrinsic relationships between grid-following and grid-forming controlled converters are established as follows: 1) the proportional gain of PLL is in inverse proportion to damping; 2) the integral gain of PLL is similar to integral droop; 3) PLL has no practical inertia but acts like grid-forming control in zero inertia cases. Further, a general stability-enhanced method combining damping and inertia is proposed, and the modified energy function is obtained to estimate the region of attraction for the system. Finally, these findings are corroborated by simulation tests with an intuitive conclusion.**

## I. INTRODUCTION

VOLTAGE source converters (VSCs) have been widely applied as an interface of renewable energy sources (RESs) to connect with the utility grid [1]-[3]. In the meanwhile, the grid-following control and grid-forming control are two types of synchronization methods for VSCs [4]. Although the grid-following control with phase-locked loop (PLL)-synchronization is the dominant control strategy for VSCs in various applications, it might suffer stability problems when the short-circuit ratio (SCR) of the grid is low [5]-[6]. To deal with this drawback, the grid-forming control like droop and virtual synchronous generator (VSG) are used to enhance the synchronization stability in weak grid cases [7].

In fact, these two control methods suffer from instability in different conditions. As for grid-following control, many numerous small-signal stability studies [8]-[9] have revealed that high bandwidth of PLL and low SCR of the grid deteriorate the small-signal stability of grid-tied VSCs. Wen et al. [8] derived impedance modeling and analyzed the influence of PLL on the impedance of VSCs. Wang et al. [9] investigated the effect of different parameters on the dominated oscillation mode associated with PLL. Meanwhile, the large-signal stability of grid-following control are analyzed in [10]-[11] and phase portraits are applied. The distance between the initial state and equilibrium point or whether there is an equilibrium point are considered as the critical factors for resynchronization of VSCs [12].

For VSCs with grid-forming control mode, the transient stability of synchronization is also of great concern [13]-[17]. In fact, the transient instability might occur even though the stable equilibrium point (SEP) exists in some cases because of the nonlinear characteristics of power transfer [5]-[6]. In [13], the modified equal-area criterion (EAC) is applied in VSG-synchronous generator (SG) systems. The current limitation is considered in [14] and the enhanced loop is added for synchronization stability. Wu et al. [15] proposed a mode-adaptive control but it might cause the oscillation of threshold. Besides, the damping effect is elaborated in [16]-[17].

To the best of our knowledge, most existing researches focus on grid-following control or grid-forming control separately. However, the intrinsic relationships between these two control schemes are rarely reported in the literature. Thus, this article attempts to fill this gap from new perspectives. Firstly, the grid-following and grid-forming control principles are established under physical models. It is found that these two synchronization frameworks are generally second-order nonlinear system. The parameter influences on synchronization stability are elaborated and the relationships between these parameters are also obtained. Moreover, the general stability-enhanced control is proposed with Lyapunov's method to analyze the transient stability of the system.

## II. THE MODELING OF CONTROLLED-SOURCES

This article aims to explore the intrinsic relationships between grid-following and grid-forming control methods. Fig. 1 illustrates the circuit and control diagrams of single-VSC infinite-bus system. $X_T$ is the transformer leakage inductance. $R_{g1,2}$, $X_{g1,2}$ are the line impedance. $v_{pcc}$ is the voltage of point of common coupling (PCC) and $v_g$ is the amplitude of grid voltage. $i_{abc}$ is the current through filter inductance $L_f$. $\omega_0$ is the nominal frequency of the system.

### A. Grid-following Control of VSCs

The grid-following method is shown in Fig. 1(a). Besides, the synchronous reference frame-based PLL (SRF-PLL) is used to synchronize with grid. The q-axis component of $v_{pcc}$ is the input of PLL and the dynamic equation of PLL is modeled as follows.

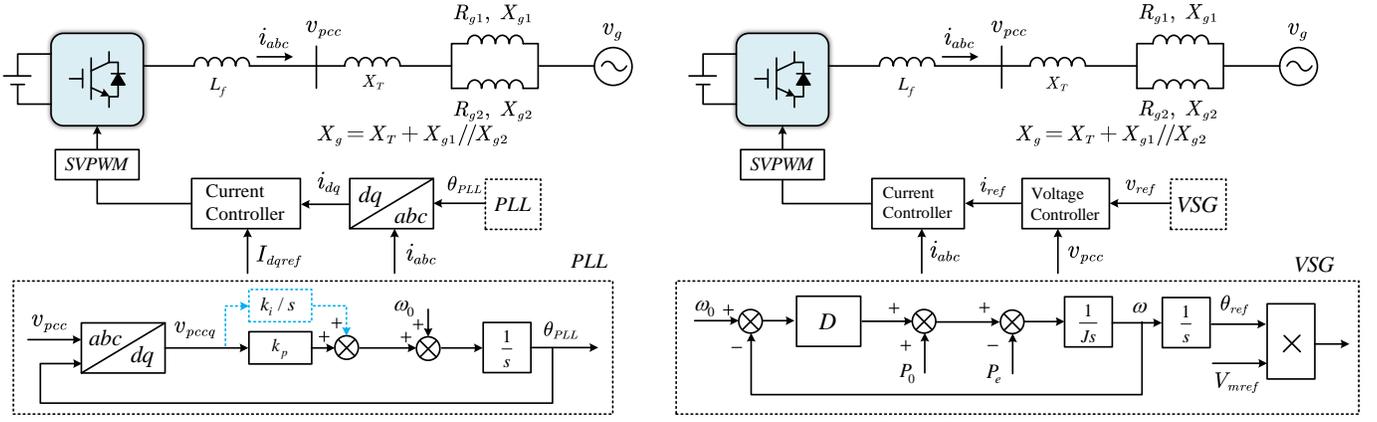

Fig. 1. Three-phase grid-connected VSC system, where its output current is within the current limit. (a) Grid- following control. (b) Grid-forming control.

$$\theta_{PLL} = \int \left[ \omega_0 + \left( k_p + k_i \int \right) v_{pccq} \right] dt \quad (1)$$

$$\delta = \theta_{pll} - \theta_g \approx \int \left( k_p + k_i \int \right) v_{pccq} dt \quad (2)$$

$$\omega_{pll} = \dot{\delta} + \omega_g \approx \dot{\delta} + \omega_0 \quad (3)$$

Where $\delta$ is the angle between $v_{pcc}$ and $v_g$. Meanwhile, the current loop can be considered as a quasi-steady-state current source in the time scale of PLL. As for transient synchronization issues, the derivative of $dq$-axis current is often ignored [9]-[11]. Thereby, neglecting the dynamics of current loop, the PCC voltage under the weak grid condition can be modeled as

$$\begin{bmatrix} v_{pccd} \\ v_{pccq} \end{bmatrix} = \begin{bmatrix} \cos\delta & \sin\delta \\ -\sin\delta & \cos\delta \end{bmatrix} \begin{bmatrix} v_g \\ 0 \end{bmatrix} + \begin{bmatrix} R_g & -\omega_{pll}L_g \\ \omega_{pll}L_g & R_g \end{bmatrix} \begin{bmatrix} i_d \\ i_q \end{bmatrix} \quad (4)$$

Where $i_{dq}$ is the $dq$-axis component of $i_{abc}$. Further, taking the derivative of $v_{pccq}$, we have

$$\dot{v}_{pccq} = -v_g \cos\delta \cdot \dot{\delta} + L_g i_d \cdot \ddot{\delta} \quad (5)$$

Combing (2)-(5), the differential equation of $\delta$ is derived and given by

$$\left( \frac{1 - k_p L_g i_d}{k_i} \right) \ddot{\delta}$$
$$= \omega_0 L_g i_d + R_g i_q - \left( \frac{k_p v_g \cos\delta}{k_i} - L_g i_d \right) \dot{\delta} - v_g \sin\delta \quad (6)$$

According to (6), the PLL-based synchronization control can be considered as the second-order swing equation. In addition, the equivalent parameters can be obtained.

$$\begin{cases} J_{eq} = \left( \dfrac{1 - k_p L_g i_d}{k_i} \right) \\ P_{0.eq} = \omega_0 L_g i_d + R_g i_q \\ D_{eq} = \left( \dfrac{k_p v_g \cos\delta}{k_i} - L_g i_d \right) \\ P_{em.eq} = v_g \end{cases} \quad (7)$$

### B. Grid-forming Control of VSCs

The grid-forming control strategy is shown in Fig. 1(b) which takes VSG as an example. The VSG controller emulates the swing equation of SG and can be expressed as [18]

$$J\ddot{\delta} = P_0 - D\dot{\delta} - P_e \quad (8)$$

Where $P_0$ and $P_e$ are the initial mechanical power and output power of VSG-based VSC. $D$ is the damping coefficient. $J$ denotes the inertia. In the inductance-dominant system, the output power of VSG is given by

$$P_e = \frac{3v_{pcc}v_g}{2X_g} \sin\delta \quad (9)$$

Substituting (8) into (7) and the second-order swing equation of VSG is obtained as

$$J\ddot{\delta} = P_0 - D\dot{\delta} - \frac{3v_{pcc}v_g}{2X_g} \sin\delta \quad (10)$$

From the analysis above, the generic model of VSCs in grid-following and grid-forming controls can be both expressed as

$$J_{eq}\ddot{\delta} = P_{0.eq} - D_{eq}\dot{\delta} - P_{em.eq} \sin\delta \quad (11)$$

## III. COMPARATIVE TRANSIENT STABILITY ANALYSIS BETWEEN GRID-FOLLOWING AND GRID-FORMING CONTROL

### A. Transient stability of VSCs in grid-following mode

In grid-following mode, the grid codes require that $i_d = 0$ and $i_q = -I_{rated}$ under grid fault to provide auxiliary voltage support [1]. Thus, Eq. (6) is modified as

$$\frac{1}{k_i}\ddot{\delta} = R_g i_q - \frac{k_p v_g \cos\delta}{k_i}\dot{\delta} - v_g \sin\delta \quad (12)$$

According to (6) and (12), the $v_{pccq} - P_{em.eq} \sin\delta$ curves are illustrated in Fig. 2 as a typical tool for stability analysis. From Fig. 2(a), the first challenge of PLL synchronization is the voltage dip and grid impedance in transient. The equilibrium point of Eq. (12) is guaranteed if there exists [14]

$$\sin\delta_s = (R_g i_q / v_g), \quad \cos\delta_s > 0 \quad (13)$$

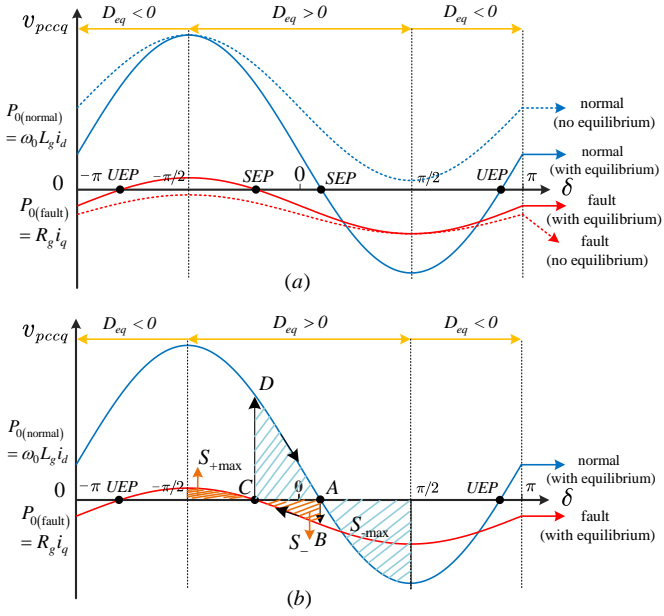

Fig. 2. The $v_{pccq} - P_{em\_eq}\sin\delta$ curves. (a) The effect of voltage dip and grid impedance. (b) The effect of second-order swing equation.

Where $\delta_s$ is a stable equilibrium point (SEP) for VSCs. However, under severe grid dip, $v_g$ might be lower than 0.2pu and (13) is unsolvable. Besides, large grid impedance increases the value of $P_{0\_eq}$ and $v_{pccq} - P_{em\_eq}\sin\delta$ curves will move away from $\delta$-axis in both normal and fault condition. These influences are described in Fig. 2(a) from solid lines to dashed lines. To enhance the synchronization stability, the frozen PLL is proposed in [19] to avoid the loss of the equilibrium point. However, it leads to static error with grid. In [10], a current injecting method is proposed to compensate the voltage drop on grid impedance and the mathematical expression is

$$\frac{i_q}{i_d} = -\frac{\omega_0 L_g}{R_g}, \; \sqrt{i_d^{\,2} + i_q^{\,2}} \le I_{rated} \qquad (14)$$

By implementing (14), $P_{0\_eq}$ is controlled equal to zero and the equilibrium point of PLL is fixed at $\delta_s = 0$. However, this kind of control methods unavoidably inject active current and violates the grid code [10]. In addition, the precise measurement of grid impedance is difficult in practical application.

Fig. 2(b) illustrates another negative factor, the second-order oscillation of (6) and (12). During the transient state of VSCs, $P_{0\_eq}$ encounters step change and PLL will move to a new equilibrium point. For instance, when stable operation point $A$ encounters grid fault and abruptly moves to $B$, the decelerating area $S$- can be defined as

$$S_- = \int_{\delta_B}^{\delta_C} v_{pccq} d\delta = \int_{\delta_B}^{\delta_C}\left(J_{eq}\ddot{\delta} + D_{eq}\dot{\delta}\right)d\delta \approx \frac{J_{eq}}{2}\dot{\delta}^2 \Big|_{\delta_B}^{\delta_C} (15)$$

When $\delta$ moves across SEP $C$, the stored kinetic energy will be released and stops $\delta$ from moving further. The accelerating area $S_+$ should not move across $\delta = -0.5\pi$ because $D_{eq}$ becomes negative. The maximum $S_+$ is given by

$$S_{+\max} = \int_{-\frac{\pi}{2}}^{\delta_C}\left(J_{eq}\ddot{\delta} + D_{eq}\dot{\delta}\right)d\delta \approx \frac{J_{eq}}{2}\dot{\delta}^2\Big|_{-\frac{\pi}{2}}^{\delta_C} \qquad (16)$$

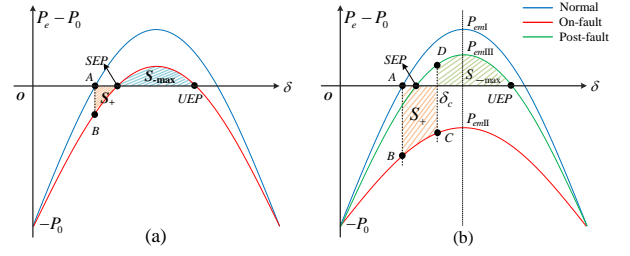

Fig. 3. The $(P_e - P_0) - \delta$ curves. (a) With equilibrium points after transient disturbance. (b) Without equilibrium point after transient disturbance.

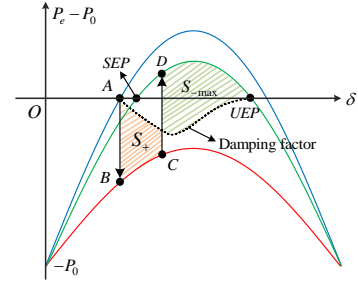

Fig. 4. The $(P_e - P_0) - \delta$ curves considering damping factor.

Hence, the sufficient condition for the resynchronization of PLL during grid fault is

$$S_{+\max} > S_- \qquad (17)$$

The similar conclusion also can be made from fault condition to normal operation. To enhance the stability of PLL during transient state, He *et al.* [11] canceled the integral part where $k_i$ of PLL is equal to 0. Thereby, the reduced model can be described as

$$\dot{\delta} = k_p\left(R_g i_q - v_g\sin\delta\right) \qquad (18)$$

Thereby, the system reduces to first-order system and can always remain stable if there exists the equilibrium point.

### B. Transient stability of VSCs in grid-forming mode

In grid-forming mode, the VSCs are controlled by VSG which emulates the swing equation of SG. Based on (10), the $(P_e - P_0) - \delta$ curves of VSG are illustrated in Fig. 3.

As seen from Fig. 3(a), the system has two equilibrium points in the first swing period, including one SEP and another unstable equilibrium point (UEP). The system might encounter this case after large disturbance like disconnection of one of the transmission lines. According to the swing equation of (10), the rotor of VSG accelerates from point $B$ until SEP due to $P_e - P_0 < 0$. When $\delta$ crosses SEP, $P_e - P_0 > 0$ and the rotor begins to decelerate till the accelerating area $S_+$ is equal to decelerating area $S$-. The critical point is UEP because $P_e - P_0 < 0$ after UEP and VSG finally loses synchronization.

Fig. 3(b) describes the system without equilibrium point after symmetrical ground fault and with equilibrium points after fault clear. The maximum electrical power of VSG under three cases is $P_{em1} > P_{em\,III} > P_{em\,II}$. The rotor of VSG keep accelerating due to $P_e - P_0 < 0$. When fault is cleared at $\delta_c$, the new equilibrium point appears and the operation point changes abruptly from $C$ to $D$. The system can maintain transient stability if $\delta$ doesn't move across UEP. If the accelerating area $S_+$ and decelerating area $S_-$ are defined as

$$\begin{cases} S_+ = \int_{\delta_n}^{\delta_c}(P_0 - P_{em\,II}\sin\delta)d\delta \approx \int_{\delta_n}^{\delta_c} J\ddot{\delta}d\delta \\ S_{-\max} = \int_{\delta_{UEP}}^{\delta_c}(P_0 - P_{em\,III}\sin\delta)d\delta \approx \int_{\delta_{UEP}}^{\delta_c} J\ddot{\delta}d\delta \end{cases} \tag{19}$$

According to EAC in power system, the stability of VSG infinite-bus system can be hold if $S_{-\max} > S_+$ [20].

From the above analysis, the damping factor is ignored for simplicity. In fact, damping is an effective way for synchronization stability and is illustrated in Fig. 4. As seen from Fig. 4, the damping decreases the mechanical power $P_0$ shown in the black dashed line, and thereby $S_+$ decreases and $S_-$ increases. This type of control methods is further improved in [13] with additional damping control and in [14] with stability enhanced $P$-$f$ droop control (SEPFC). In addition, a typical Lyapunov function is established in [20] to prove the stability of damping control.

Meanwhile, unlike the constant inertia in conventional SG, the virtual inertia can be adaptively tuned based on the proper algorithm. In fact, $\delta$ is the function of time. Based on (10), the calculation of $\delta$ from $t_n$ to $t_{n+1}$ is given by

$$\Delta\delta_{(n+1)} = \Delta\delta_{(n)} + \frac{(P_0 - P_{e(n)})}{J}\Delta t^2 = \Delta\delta_{(n)} + a_{(n)}\Delta t^2 \tag{20}$$

$$\delta_{(n+1)} = \delta_{(n)} + \Delta\delta_{(n+1)} \tag{21}$$

Where $a_{(n)}$ is the acceleration of rotor and it is in inverse proportion to inertia $J$. Thereby, the inertia is tuned large in deviation stage and small in returning stage. The mathematical expression of adaptive inertia is given as follows [18].

$$J = J_0 + k(\omega - \omega_0)\frac{d\omega}{dt} \tag{22}$$

Where $J_0$ is the constant inertia and $k$ is the adaptive gain.

### C. Analogy between grid-following and grid-forming control

Comparing (6) with (10), the second-order oscillation is generated by $k_i$ in PLL and $J$ in swing equation. Therefore, by setting $k_i = 0$ and $J = 0$, we have

$$\begin{cases} \dot{\delta}_{pll} = k_p\left(R_g i_q - v_g\sin\delta\right) \\ \dot{\delta}_{VSG} = \dfrac{1}{D}\left(P_0 - \dfrac{3v_{pcc}v_g}{2X_g}\sin\delta\right) \end{cases} \tag{23}$$

From (23), the systems in grid-following and grid-forming control are both reduced to first-order system like droop control and can remain stable with existence of equilibrium point. It should be noted that the integral coefficient $k_i$ isn't physically equal to inertia $J$. More specifically, $k_i$ is similar to integral droop coefficient $m_i$ proposed in our previous work to eliminate the static error of output power in islanded mode [3].

In addition, the large increase of $k_i$ and $m_i$ will generate instable poles [3], [8].

Further, setting $k_p = 0$ and PLL is frozen in grid-following control [19]. It is equivalent to set infinite damping and zero droop coefficient in grid-forming control. Thereby, the synchronization stability of VSCs is achieved, but the output power of VSCs is out of control. Therefore, $k_p$ is in inverse proportion to damping $D$. Small value of $k_p$ means stable operation and large value of $k_p$ improves PLL's dynamics.

In general, the relationships among these two synchronization methods are summarized in Table I.

## IV. TRANSIENT STABILITY IMPROVEMENT

### A. Stability-enhanced control combing damping and inertia

Based on the analysis in Section III, the synchronization stability of VSCs is affected by the damping and inertia coefficient in swing equation. In this article, the general stability-enhanced control is presented as follows by combing extra damping control and adaptive inertia.

$$J_{ad}\ddot{\delta} = P_0 - (D + k_\omega)\dot{\delta} - P_{em}\sin\delta \tag{24}$$

$$J_{ad} = \begin{cases} nJ_0\,; & if\,(\omega - \omega_0)\cdot\dfrac{d\omega}{dt} > 0 \\ J_0\,; & else \end{cases} \tag{25}$$

Where $k_\omega$ is the speed governor gain to enhance the damping. $J_{ad}$ is the adaptive inertia and $n > 1$. When $\dot{\delta}$ and $\ddot{\delta}$ are in the same direction, the inertia increases to reduce the deviation of $\omega$. Otherwise, the inertia returns to $J_0$ to accelerate dynamics and improve the stability of the system.

### B. Implementation of Stability-enhanced control for grid-following control

For VSCs in grid-following control, they don't have practical inertia. During the transient, the integral coefficient $k_i$ is advised to set zero to avoid oscillation of PLL.

As for the design of proportional coefficient $k_p$, the adaptive tuning by $v_{pccq}$ under grid fault is proposed in this paper and the mathematical expression of $k_p$ is given by

$$k_{p.ad} = \begin{cases} 0 & ;\,v_{pccq}\in\left[-\infty, -\dfrac{k_p}{k_{vq}}\right] \\ k_p + k_{vq}v_{pccq} & ;\,v_{pccq}\in\left[-\dfrac{k_p}{k_{vq}}, 0\right] \\ k_p & ;\,v_{pccq}\in[0,\infty] \end{cases} \tag{26}$$

Where $k_{p.ad}$ is the adaptive proportional gain of PLL and $k_{vq}$ is the stability-enhanced coefficient. During the transient of grid dip, larger $k_p$ leads to more shift from A to B as shown in Fig. 2(b) and the area of $S_-$ increases, causing less stability margin. After implementing adaptive $k_{p.ad}$ in Eq. (26), the output frequency of PLL becomes

TABLE I. ANALOGOUS RELATIONSHIPS BETWEEN TWO TYPES OF SYNCHRONIZATION METHODS

| Grid-following Control | Grid-forming Control | Relationships and Influences |
|---|---|---|
| $k_p$ | $D$ | $k_p \propto (1/D)$, similar to frequency droop gain and in inverse proportion to damping |
| $k_i$ | $J$ | $k_i = J = 0$, reducing to first-order system and avoid oscillation |
| $k_i$ | $m_i$(grid − following) | $k_i \propto m_i$, integral coefficient to eliminate the static error of output power |
| $\omega_c$(grid − forming) | $J$ | $J \propto (1/\omega_c)$, physic inertia and inverse proportion to cut-off frequency of power filter |

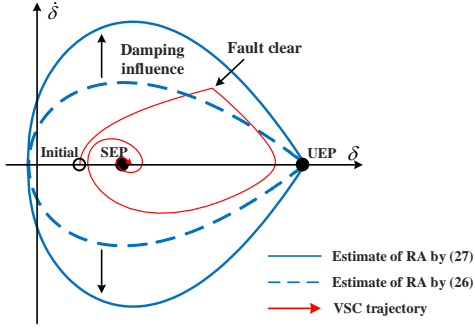

Fig. 5. Estimate of RA and VSC trajectory.

$$\omega_{pll} = \omega_0 + k_{p\_ad}v_{pccq} = \omega_0 + k_p v_{pccq} + \underbrace{k_{vq}v_{pccq}^2}_{\text{quadratic}} \quad (27)$$

Thereby, the quadratic term of (27) decreases the area of $S_-$ and helps maintain the synchronization stability of PLL. Moreover, when there's no equilibrium point, the divergency of PLL can be avoided, since $k_{p.ad}$ is adaptively tuned to zero and PLL is frozen. After grid fault is cleared, $v_{pccq}$ becomes positive and $k_{p.ad}$ switches to ordinary PLL.

From the analysis above, it can be deduced that the proposed adaptive $k_{p.ad}$ can be regarded as a nonlinear controller, as $v_{pccq}$ is a nonlinear function of $\delta$. In addition, it can avoid synchronization instability under severe grid fault and can switch to ordinary PLL softly when fault is cleared.

### C. Stability analysis through Lyapunov function

Based on the generic model in (11), the energy function of single-VSC infinite-bus is derived as follows [4], [20].

$$V(\delta, \dot{\delta}) = \frac{J\dot{\delta}^2}{2} + (E_0 - P_0\delta - P_{em}\cos\delta) = E_k + E_p \quad (28)$$

Where $E_k$ and $E_p$ are the equivalent kinetic energy and potential energy. However, (28) doesn't consider the damping and estimate of the region of attraction (RA) is conservative. In this article, the modified energy function is proposed as

$$V(\delta, \dot{\delta}) = \frac{J\dot{\delta}^2}{2} + (E_0 - P_0\delta - P_{em}\cos\delta) + D|\delta\dot{\delta}| \quad (29)$$

Further, if there exist equilibrium points for VSCs, the derivative of $V(\delta, \dot{\delta})$ can be derived as follows.

$$\dot{V}(\delta, \dot{\delta}) = -\frac{D}{J}|\delta(P_0 - D\dot{\delta} - P_{em}\sin\delta)| \quad (30)$$

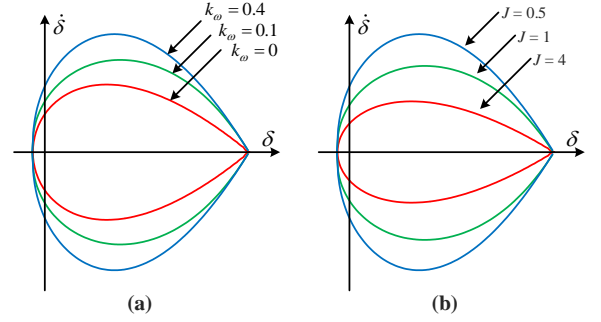

Fig. 6. Estimate of RA. (a) With variation of $k_\omega$. b) With variation of $J$.

According to the Local La Salle's Invariance Principle, the global $V(\delta, \dot{\delta}) > 0$ is not satisfied, but the local stability of the system can be obtained [21]. Therefore, the estimate of RA is implemented by

$$\Omega_c = \left\{ V(\delta, \dot{\delta}) < c, c > 0, \dot{V}(\delta, \dot{\delta}) \le 0 \right\} \quad (31)$$

Usually, $c$ is calculated by the local UEP. The estimate of RA is shown in Fig. 5 with parameters detailed in Appendix. From Fig. 5, the modified energy function in (29) can estimate RA more accurately with consideration of damping.

Fig. 6 illustrates the influence of extra damping gain $k_\omega$ and inertia $J$, respectively. The increase of damping improves the stability of the system. In contrast, large inertia decreases RA. Thereby, the adaptive inertia needs to return to $J_0$ in decelerating stage after fault clear.

## V. SIMULATION VERIFICATION

In this section, the detailed single-VSC infinite-bus system is built in MATLAB/Simulink platform. The topology is shown as Fig. 1 and key parameters are given in Appendix.

First, the grid-tied VSC in grid-following control is tested and the results are illustrated in Fig. 7. The performance with original PLL control is given in Fig. 7(a). The grid dip happens at 2s and VSC loses synchronization with grid. When fault is cleared at 3s, VSC achieves resynchronization. In contrast, VSC with stability-enhanced control in Fig. 7(b) remains stable both in grid dip and postfault conditions. Thereby, the reduced first-order VSC with smaller PLL gain $k_p$ is more robust to large disturbances of grid.

Second, the grid-tied VSC in grid-forming control is implemented, where the current limit is only activated during grid fault. The grid fault occurs at 2s and is cleared at 2.4s. As seen from Fig. 8(a), under original VSG control, the rotor accelerates fast and loses synchronization with grid after grid fault. Besides, the VSC can't resynchronize with grid after first cycle of oscillation due to the lack of damping. Fig. 8(b) illustrates the performance of the system with stability-enhanced VSG control. When grid fault occurs at 2s, the adaptive inertia and damping prevent rotor to accelerate. When fault is cleared at 2.4s, the inertia decreases to $J_0$ for fast returning. Thereby, the stability of the system is enhanced.

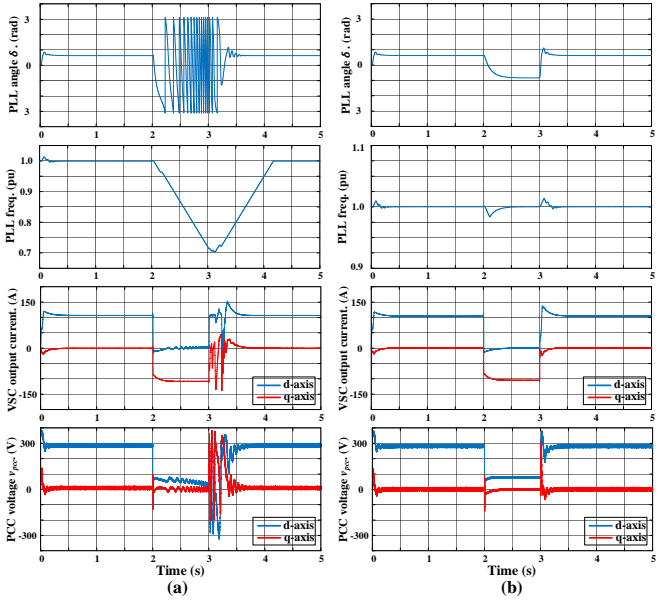

Fig. 7. Results for grid-following VSC. (a) With original PLL. b) With stability-enhanced control of PLL.

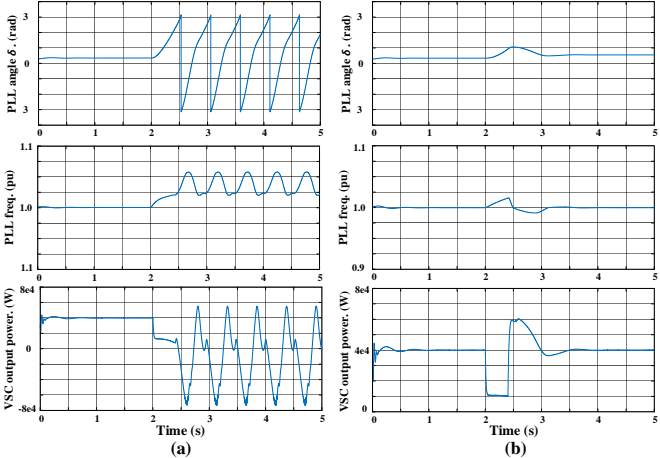

Fig. 8. Results for grid-forming VSC. (a) With original VSG. b) With stability-enhanced VSG control.

## VI. CONCLUSION

In this paper, the intrinsic relationships and differences among grid-following and grid forming controls are explored. Based on the theoretical analysis, the generic second-order models are established. The proportional coefficient of PLL is in inverse proportion to damping, and the integral coefficient of PLL is similar to integral droop rather than the inertia of SG. Further, the general stability-enhanced control is proposed with extra damping and adaptive inertia control. The modified energy function is obtained to quantify the transient stability of the system. Based on these findings, future efforts will be devoted to the research on the transient stability of more complex systems.

## APPENDIX

The parameters of studied systems are given as follows:

TABLE II
KEY PARAMETERS OF TESTED SYSTEMS

| The base of the per-unit system: |
|---|
| $S_n = 0.05MVA$, $V_n(\text{pk}) = 311V$, $\omega_0 = 100\pi\ rad/s$, $I_m = 107A$; |
| $X_f = 0.15pu$, $X_T = 0.2pu$, $X_{g1} = X_{g2} = 0.5pu$, $R_{g1} = R_{g1} = 0.05pu$, $V_{mref} = 1.05pu$, $v_g = 1\ or\ 0.2pu$; |
| Grid-following control: |
| $i_d = 1\ or\ 0pu$, $i_q = 0\ or\ -1pu$, |
| $k_p + k_i/s = 0.3 + 4/s$; |
| Grid-forming control: |
| $P_{0,eq} = 0.8pu$, $J_{eq} = 300$, $n = 5$, $D_{eq} + k_\omega = 4e3$; |
| Voltage loop control: $0.5 + 50/s$; |
| Current loop control: $50 + 500/s$. |

## REFERENCES

[1] M. G. Taul, X. Wang, P. Davari, and F. Blaabjerg, "An overview of assessment methods for synchronization stability of grid-connected converters under severe symmetrical grid faults," in IEEE Transactions on Power Electronics, vol. 34, no. 10, pp. 9655–9670, 2019.

[2] L. Huang, H. Xin and F. Dörfler, "H ∞ -Control of Grid-Connected Converters: Design, Objectives and Decentralized Stability Certificates," in IEEE Transactions on Smart Grid, vol. 11, no. 5, pp. 3805-3816, Sept. 2020.

[3] M. Su, K. Zhang, Y. Sun, H. Han, G. Shi, S. Fu, "Coordinated control for PV-ESS islanded microgrid without communication," International Journal of Electrical Power & Energy Systems, vol. 136, Oct. 2021.

[4] X. Fu et al., "Large-Signal Stability of Grid-Forming and Grid-Following Controls in Voltage Source Converter: A Comparative Study," in IEEE Transactions on Power Electronics, vol. 36, no. 7, pp. 7832-7840, July 2021, doi: 10.1109/TPEL.2020.3047480.

[5] H. Wu and X. Wang, "Design-oriented transient stability analysis of grid-connected converters with power synchronization control," in IEEE Transactions on Industry Electronics, vol. 66, no. 8, pp. 6473–6482, Aug. 2019.

[6] W. Dong, H. Xin, D. Wu and L. Huang, "Small Signal Stability Analysis of Multi-Infeed Power Electronic Systems Based on Grid Strength Assessment," in IEEE Transactions on Power Systems, vol. 34, no. 2, pp. 1393-1403, March 2019, doi: 10.1109/TPWRS.2018.2875305.

[7] Z. Shuai, C. Shen, X. Liu, Z. Li, and Z. J. Shen, "Transient angle stability of virtual synchronous generators using Lyapunov's direct method," in IEEE Transactions on Smart Grid, vol. 10, no. 4, pp. 4648–4661, Jul. 2019.

[8] B. Wen, D. Boroyevich, R. Burgos, P. Mattavelli, and Z. Shen, "Analysis of D-Q small-signal impedance of grid-tied inverters," in IEEE Transactions on Power Electronics, vol. 31, no. 1, pp. 675–687, Jan. 2016.

[9] X. Wang, J. Yao, J. Pei, P. Sun, H. Zhang, and R. Liu, "Analysis and damping control of small-signal oscillations for VSC connected to weak AC grid during LVRT," in IEEE Transactions on Energy Conversion, vol. 34, no. 3, pp. 1667–1676, Sep. 2019.

[10] S. Ma, H. Geng, L. Liu, G. Yang, and B. C. Pal, "Grid-synchronization stability improvement of large scale wind farm during severe grid fault," in IEEE Transactions on Power Systems, vol. 33, no. 1, pp. 216–226, Jan. 2018.

[11] X. He, H. Geng, J. Xi and J. M. Guerrero, "Resynchronization Analysis and Improvement of Grid-Connected VSCs During Grid Faults," in IEEE Journal of Emerging and Selected Topics in Power Electronics, vol. 9, no. 1, pp. 438-450, Feb. 2021.

[12] C. Wu, X. Xiong, M. G. Taul, and F. Blaabjerg, "Enhancing transient stability of pll-synchronized converters by introducing voltage

normalization control," in IEEE Journal on Emerging and Selected Topics in Circuits and Systems, vol. 11, no. 1, pp. 69–78, 2021.

[13] H. Cheng, Z. Shuai, C. Shen, X. Liu, Z. Li and Z. J. Shen, "Transient Angle Stability of Paralleled Synchronous and Virtual Synchronous Generators in Islanded Microgrids," in IEEE Transactions on Power Electronics, vol. 35, no. 8, pp. 8751-8765, Aug. 2020.

[14] L. Huang, H. Xin, Z. Wang, L. Zhang, K. Wu, and J. Hu, "Transient stability analysis and control design of droop-controlled voltage source converters considering current limitation," in IEEE Trans. Smart Grid, vol. 10, no. 1, pp. 578–591, Jan. 2019.

[15] H. Wu and X. Wang, "A Mode-Adaptive Power-Angle Control Method for Transient Stability Enhancement of Virtual Synchronous Generators," in IEEE Journal of Emerging and Selected Topics in Power Electronics, vol. 8, no. 2, pp. 1034-1049, June 2020.

[16] L. Huang, H. Xin, H. Yuan, G. Wang, and P. Ju, "Damping effect of virtual synchronous machines provided by a dynamical virtual impedance," in IEEE Transactions on Energy Conversion., vol. 36, no. 1, pp. 570–573, Mar. 2021.

[17] C. Yang, L. Huang, H. Xin and P. Ju, "Placing Grid-Forming Converters to Enhance Small Signal Stability of PLL-Integrated Power Systems," in IEEE Transactions on Power Systems, vol. 36, no. 4, pp. 3563-3573, July 2021.

[18] X. Hou, Y. Sun, X. Zhang, J. Lu, P. Wang and J. M. Guerrero, "Improvement of Frequency Regulation in VSG-Based AC Microgrid Via Adaptive Virtual Inertia," in IEEE Transactions on Power Electronics, vol. 35, no. 2, pp. 1589-1602, Feb. 2020.

[19] B. Weise, "Impact of k-factor and active current reduction during fault-ride-through of generating units connected via voltage-sourced converters on power system stability," IET Renewable Power Generation, vol. 9, no. 1, pp. 25–36, 2015.

[20] P. Kundur, Power System Stability and Control. New York, NY, USA: McGraw-Hill, 1994.

[21] H. K. Khalil, Nonlinear Systems. New Jersey, NJ, USA: Prentice Hall, 2002, pp. 111–116.